\def\be{\begin{equation}}
\def\ee{\end{equation}}
\def\A{\vec{A}}
\def\B{\vec{B}}
\def\D{\nabla}
\def\d{\delta^3(\vec{x}-\vec{y})}
\def\E{\vec{E}}

\def\nn{\nonumber}

\def\ni{\noindent}

\def\XL{\tilde{X}^L}
\def\XR{\tilde{X}^R}

\def\tG{\tilde{G}}

\def \overpmup {\overleftarrow{\partial}\,\,\!\!\!\!\!\!\!
\overrightarrow{\hbox{\ \ }^\mu}}

\documentstyle[12pt,a4]{article}

\begin{document}
\thispagestyle{empty}
\vskip20mm
\begin{center}

{\bf THE ELECTROMAGNETIC AND PROCA FIELDS REVISITED:\\
A UNIFIED QUANTIZATION} 
\footnote[1]{Work partially supported by the Comisi\'on
Interministerial de Ciencia y Tecnolog\'\i a.}
\vskip7mm
{\it V\'\i ctor Aldaya$^{1,2}$, Manuel Calixto$^{1,5}$   
and Miguel Navarro$^{3,4}$ } 
\vskip 5mm
\end{center}
\footnotesize
\begin{enumerate}
\item Instituto Carlos I de F\'\i sica Te\'orica y Computacional,
Facultad  de  Ciencias, Universidad de Granada, Campus de Fuentenueva,
18002, Granada, Spain.
\item IFIC, Centro Mixto Universidad de
Valencia-CSIC, Burjassot 46100-Valencia, Spain.
\item The Blackett Laboratory, Imperial College, Prince 
Consort Road, London SW7 2BZ; United Kingdom.
\item Instituto de Matem\'aticas y F\'\i sica Fundamental, CSIC, Serrano
113-123, 28006 Madrid, Spain
\item Departamento de F\'\i sica Te\'orica y del Cosmos, Facultad
de Ciencias, Universidad de Granada, Campus de Fuentenueva, 
Granada 18002, Spain
\vskip10mm
\normalsize
\end{enumerate}
\centerline{\bf Abstract}
\vskip5mm
\footnotesize
Quantizing the electromagnetic field with a group 
formalism faces the difficulty 
of how to turn the traditional gauge transformation 
of the  vector potential, 
$A_{\mu}(x)\rightarrow A_{\mu}(x)+\partial_{\mu}\varphi(x)$, 
into a group law.  
In this paper it is shown that the problem can 
be solved by looking at gauge 
transformations in a slightly different manner
which, in addition, does not require introducing 
any BRST-like parameter. This gauge transformation does not appear explicitly 
in the 
group law of the symmetry but rather as the trajectories associated with 
generalized equations of motion generated by vector fields with null 
Noether invariants. In the new approach the parameters of the local 
group, $U(1)(\vec{x},t)$, acquire dynamical content outside the photon 
mass shell, 
a fact which also allows a unified quantization of both 
the electromagnetic and Proca fields. 
\normalsize


\vfil\eject

\section{Introduction}
According to the minimal interaction principle, 
in order to extend any internal
symmetry of the matter fields to the gauge level, i.e. turning the group 
parameters $\varphi^a$ into functions on configuration space, a vector 
potential
$A_{\mu}^a(x)$ which 
transforms as a connection form must be introduced . For the
particular case of the $U(1)$-symmetry 
this vector potential acquires only a
derivative of the gauge parameter under a gauge transformation: 

\begin{equation}
A_{\mu}(x)\rightarrow A_{\mu}(x)+
\partial_{\mu} \varphi(x) \; .\label{gauge}
\end{equation}

\ni This is the standard way of introducing the electromagnetic 
interaction in a geometric formulation. However, the electromagnetic field has 
its own entity, not necessarily attached to the non-tensorial part of the 
gauge transformations, and its quantization can be achieved directly in terms 
of the field strengths \cite{Birula}. Quantizing in terms of $A_{\mu}$ results
in a redundant system which must be further constrained by the so-called 
{\it Gauss law}. Let us briefly review the quantum origin of this constraint. 

Maxwell's theory for the electromagnetic field without sources may be
derived from the Lagrangian density: \be {\cal
L}=-\frac{1}{4}F^{\mu\nu}F_{\mu\nu}=\frac{1}{2} (\vec{E}^2-\vec{B}^2) \; ,
\label{lag} \ee \ni where \be F_{\mu\nu}\equiv\partial_\mu A_\nu
-\partial_\nu A_\mu,\;\;\; \vec{E}\equiv - \dot{\vec{A}}-\vec{\nabla}
A^0,\;\;\; \B\equiv\D\times\A \ee \ni are the electromagnetic tensor and
the electric and magnetic fields, respectively. The equations of motion
are: \be \partial_\mu F^{\mu\nu}=0 \; , \label{eq1} \ee \ni for which the
time and space components are the Gauss law and the Amp\`ere law,
respectively: \be \D\cdot\E=0, \;\;\;\;\; \dot{\E}=\D\times\B \; . \ee \ni
The canonical quantization of this system encounters a problem when
submitted to a Hamiltonian description, due to the fact that the
Lagrangian (\ref{lag}) does not depend on $\dot{A}^0$, so that there is no
momentum conjugate to $A^0$.  In other words, this Lagrangian is not
regular.  This means, in general, that not all Lagrangian equations can be
written in Hamiltonian form and some must be added to the set of Hamilton
equations as constraints. This is precisely the case of the Gauss law when
one uses the gauge freedom to set the non-covariant condition $A^0=0$,
i.e. the so-called ``Weyl gauge" (see e.g. \cite{Jackiw}). 

In a previous paper \cite{newlook} the quantization 
of the free electromagnetic field was achieved in such a way 
that the Gauss law appeared on the same footing
as the ordinary Hamilton equations of motion 
and not as a constraint. The 
algorithm used was a group approach to 
quantization (GAQ) \cite{23,Ramirez} formulated 
on an infinite-dimensional Lie group 
parameterized by the strengths
$\vec{E}(x)$ and $\vec{B}(x)$, as well as the time variable. 
However, the 
analogous quantization of the free electromagnetic 
system in terms of the 
vector potential $A_{\mu}$ found it difficult to 
accommodate the gauge 
transformation property (\ref{gauge}) to a group law. 
The introduction of an extra group parameter,
which revealed itself as being some sort of bosonic 
BRST parameter \cite{BRST}, was apparently needed. 

In this paper we propose a new infinite-dimensional 
Lie group $\tG$, with a 
principal bundle structure $\tG\rightarrow\tG/T$, parameterized, 
roughly speaking, 
by $A_{\mu}(\vec{x},t)$, the Poincar\'e variables and the coordinates of the 
local $U(1)(\vec{x},t)$, which plays the role 
of the structure 
group $T$. This subgroup generalizes the standard $U(1)$-phase 
invariance in Quantum Mechanics, so that the extra
equivariance conditions associated 
with the local subgroup $U(1)(\vec{x},t)$ will 
provide the traditional constraints of the theory. 
The construction of the new group law has required a review of the concepts 
of gauge symmetry and constraints and has led, as a byproduct, 
to a unified quantization of both the electromagnetic and Proca fields, 
within the same general scheme of 
quantization based on a group (GAQ). 

Let us motivate the explicit group law to be proposed. Going back 
to the Lagrangian analysis of the origin of the 
constraints outlined above, we
must note that the non-regularity property of the 
Lagrangian (\ref{lag}) can be
covariantly expressed by stating that ${\cal L}$ does 
not depend on the covariant 
quantity $\Phi\equiv\partial_\mu A^\mu$. This suggests resorting to the new, 
regular Lagrangian (the Fermi Lagrangian if $\lambda=1$)

\be
{\cal L}'=-\frac{1}{4}F^{\mu\nu}F_{\mu\nu}-\frac{\lambda}{2}
(\partial_\mu A^\mu)^2,\label{lag2}
\ee

\ni for arbitrary $\lambda$, which keeps a residual 
(covariant) gauge invariance 
$A_{\mu}(x)\rightarrow A_{\mu}(x)+
\partial_{\mu} \varphi(x)$ under (on-shell) functions
$\varphi(x)$ such that $\partial_\mu\partial^\mu\varphi(x)=0$. 
With this residual symmetry, we could 
formally associate a Noether charge of the form

\be
Q_\varphi=\int{d^3x\frac{\partial{\cal L}'}{\partial \dot{A}_\mu}\partial_\mu
\varphi}=\lambda\int{d^3x(\dot{\Phi}\varphi-\Phi\dot{\varphi})}, \label{carga}
\ee

\ni where we have used the equations of motion:

\be
\partial_\mu F^{\mu\nu}+\lambda\partial^\nu\Phi=0\Rightarrow 
\partial_\mu\partial^\mu\Phi=0 .
\label{eq2}
\ee

\ni The standard Maxwell's equations (\ref{eq1}) can be regained simply by  
putting $\Phi=\hbox{constant}$ as a (classical) constraint. In particular, 
the constraint $\dot{\Phi}=0$ in (\ref{eq2}) reproduces the Gauss Law.

A glance at eq. (\ref{carga}) reveals that the quantity $\dot{\Phi}$ behaves 
as a generator of gauge transformations (those depending only on the 
Cauchy hypersurface parameters; see \cite{Jackiw} for the 
non-covariant case). In fact, both $\Phi$ and $\dot{\Phi}$ 
close a Poisson algebra with the rest of the dynamical variables and 
the Hamiltonian associated with ${\cal L}'$. 
It is straightforward to compute the 
following Poisson brackets ($\lambda=1$
\footnote{We adopt this choice for simplicity, but the  
results are $\lambda$ independent.}):

\be
\begin{array}{ll}\{{A}_\mu(\vec{x},t),{A}_\nu(\vec{y},t)\}=0 & 
\{\dot{A}_\mu(\vec{x},t),\dot{A}_\nu(\vec{y},t)\}=0 \\
\{\dot{A}_\mu(\vec{x},t),A^\nu(\vec{y},t)\}=g^\nu_\mu\d &
\{\dot{\Phi}(\vec{x},t),\Phi(\vec{y},t)\}=0 \\
\{\Phi(\vec{x},t),A^\mu(\vec{y},t)\}=g^\mu_0\d &
\{\Phi(\vec{x},t),\dot{A}^\mu(\vec{y},t)\}=-g^\mu_i\partial^i\d \\
\{\dot{\Phi}(\vec{x},t),A^\mu(\vec{y},t)\}=g^\mu_i\partial^i\d &
\{\dot{\Phi}(\vec{x},t),\dot{A}^\mu(\vec{y},t)\}=-g^\mu_0\D^2\d \\
\{A_\mu,H\}=\dot{A}_\mu & \{\dot{A}_\mu,H\}=\D^2 A_\mu \\ 
\{\Phi,H\}=\dot{\Phi} & \{\dot{\Phi},H\}=\D^2\Phi 
.\end{array}\label{poisson}
\ee

\ni The statement that the Maxwell's tensor is gauge invariant, can now be  
expressed by:  $\{F_{\mu\nu}(\vec{x},t),\Phi(\vec{y},t)\}=
\{F_{\mu\nu}(\vec{x},t),\dot{\Phi}(\vec{y},t)\}=0$, as it can be easily
derived from (\ref{poisson}). 

As a Lie algebra, (\ref{poisson}) is a central 
extension by $U(1)$ characterized
by a (Lie algebra) co-cycle involving the generators 
$(A,\dot{A},\Phi,\dot{\Phi})$, much in the same way the Poisson bracket
$\{p,q\}=1$ characterizes the (Lie algebra) co-cycle of the Heisenberg-Weyl 
subalgebra for a particle in particle Mechanics. From this co-cycle we can read 
immediately that the couple $(A,\dot{A})$ 
corresponds to a canonically-conjugate 
pair of variables and that the $(\Phi,\dot{\Phi})$ variables are not completely 
devoid of dynamical content, as the bracket 
$\{\dot{\Phi}(\vec{x},t),\Phi(\vec{y},t)\}=0$ might 
suggest at first sight, due to the {\it non-diagonal terms in the co-cycle} 
(lines $3^{\hbox{\small rd}}$ and $4^{\hbox{\small th}}$ in (\ref{poisson})). 
For this reason, the variables $\Phi,\dot{\Phi}$ cannot be properly 
associated with the usual null Noether charge gauge generators. 
Rather, these variables will be considered simply as 
the generators of the structure group $T$ and related with the constraints.

As our starting point to perform 
the quantization of the electromagnetic and Proca fields,  
we shall adopt the abstract structure of the Poisson algebra above (with the 
Poincar\'e generators added), but we shall 
deform it with a non-trivial central term of the form
$\{\dot{\Phi}(\vec{x},t),\Phi(\vec{y},t)\}=m^2\d$ ($m$ being a parameter with 
mass dimension; we are using natural 
unities, $\hbar=1=c$) and construct the group law by 
standard exponentiation. The new 
central term, parameterized by $m$, modifies the dynamical content of the 
group variables, transferring degrees of freedom between the $A$ and $\Phi$ 
variables. In the case $m\not= 0$ the group co-cycle diagonalizes in a new 
set of variables which correspond to the 
Proca field and some sort of scalar field.

\section{Unified quantization 
of the electromagnetic and Proca fields}

According to the general prescription of GAQ we shall 
start from a group law, which is inspired in the 
(classical) Poisson algebra (\ref{poisson}), 
involving the Poincar\'e parameters 
$x^{\mu},\Lambda^{\mu}_{\nu}$, the Fourier
coefficients $a_{\mu}(k),\;a^+_{\mu}(k)$ of the field $A_{\mu}$, 
the Fourier coefficients $\phi(k),\;\phi^+(k)$ 
of the local $U(1)_{loc}$ subgroup 
and the parameter $\zeta$ associated with 
the central $U(1)$ subgroup. The entire
group $\tilde{G}$ will be regarded either as: a) 
a non-central extension by a 
group $T$ (parameterized by $\phi(k),\;\phi^+(k)$ and $\zeta$) of the group 
constituted by the space-time symmetry 
and the Fourier coefficients of $A_{\mu}$, or b)
a central extension by $U(1)$ of the remainder: $G\equiv\tG/U(1)$. 
In both cases the group law
corresponding to the central parameter will 
be characterized by a set of cocycles $\xi_1,\;\xi_2,\;\xi_3$ (see below)
defined on the whole ``classical" group $G$. 
As mentioned above, the equations associated
with the $T$-equivariance condition, 
other than the usual $U(1)$ condition, are
interpreted as constraints in the general group-quantization 
formalism \cite{Ramirez}.

The precise group law $g''=g'*g$ is:

\begin{eqnarray}
a_{\mu}''(k)&=&a_{\mu}'(k)\exp(-ik\cdot\Lambda'x)+
{\Lambda'}^{\nu}_{\mu}a_{\nu}(\Lambda^{-1}{}'k) \nn \\
a_{\mu}^+{}''(k)&=&a_{\mu}^+{}'(k)\exp(ik\cdot\Lambda'x)+
{\Lambda'}^{\nu}_{\mu}a_{\mu}^+(\Lambda^{-1}{}'k) \nn \\ 
\phi''(k)&=&\phi'(k)\exp(-ik\cdot\Lambda'x)+
\phi(\Lambda^{-1}{}'k) \nn \\
\phi^+{}''(k)&=&\phi^+{}'(k)\exp(ik\cdot\Lambda'x)+
\phi^+(\Lambda^{-1}{}'k) \nn \\ 
x''&=&x'+\Lambda'x \nn \\ 
\Lambda''&=&\Lambda'\Lambda \label{law} \\
\zeta''&=&\zeta'\zeta \exp\left[\frac{i}{2}\int\frac{d^3k}{2k^0}
i\left\{\xi_1(g',g)+\xi_2(g',g)+\xi_3(g',g)\right\}\right], 
\;\;\;g,g'\in G \nn \\ 
\xi_1(g',g)&\equiv& g^{\mu\sigma}\Lambda^{\nu}_{\sigma}{}'
[a_{\mu}(\Lambda^{-1}{}'k)a_{\nu}^+{}'(k)e^{ik\cdot\Lambda'x}- 
a_{\mu}^+(\Lambda^{-1}{}'k)a_{\nu}'(k)e^{-ik\cdot\Lambda'x}] \nn \\
\xi_2(g',g)&\equiv& i[\phi^+{}'(k)(\Lambda^{-1}{}')^{\mu\nu}k_{\nu}a_{\mu}
(\Lambda^{-1}{}'k)e^{ik\Lambda'x}+ 
\phi'(k)(\Lambda^{-1}{}')^{\mu\nu}k_{\nu}a^+_{\mu}
(\Lambda^{-1}{}'k)e^{-ik\Lambda'x}\nn \\
& &-k^{\mu}a^+_{\mu}{}'(k)\phi
(\Lambda^{-1}{}'k)e^{ik\cdot\Lambda'x} 
- k^{\mu}a_{\mu}'(k)\phi^+(\Lambda^{-1}{}'k)
e^{-ik\cdot\Lambda'x}]\nn \\
\xi_3(g',g)&\equiv&k^2[\phi^+{}'(k)\phi(\Lambda^{-1}{}'k)e^{ik\cdot\Lambda'x} 
 - \phi'(k)\phi^+(\Lambda^{-1}{}'k)e^{-ik\cdot\Lambda'x}]. \nn
\end{eqnarray}

Note that, in fact, we are not actually dealing with the whole local group 
$U(1)_{loc}$ but rather with a subgroup made of ``on-shell" functions and 
related to the residual gauge invariance surviving in the Lagrangian 
(\ref{lag2}); nevertheless, we shall refer to this symmetry simply as local 
$U(1)_{loc}$ symmetry. A remarkable feature of the proposed group 
law is that, unlike 
in the standard formulation of gauge theories, the local 
$U(1)_{loc}$ subgroup does not act on the vector potential parameters 
(as in Eq. (\ref{gauge})), thus obviating the need of introducing new 
(BRST-like) parameters as in Ref. \cite{BRST}. 
However,  the local symmetry acts 
non-trivially on the $U(1)$ parameter $\zeta$, causing a change in the 
phase of the wave function (see below). Furthermore, the traditional 
transformation properties of the vector 
potential will appear as the trajectories 
of some of the generalized (quantum) equations of motion. 

The arbitrariness in the choice of the value of $k^2$ 
in the co-cycle $\xi_3(g',g)$, 
extending the local $U(1)_{loc}$ subgroup, will lead 
to a unified quantization of the electromagnetic field ($k^2=0$ case) and 
the Proca field ($k^2=m^2\not=0$ case) 
when the $T$-equivariant conditions are imposed 
as constraints in both cases. We shall 
maintain $k^2$ throughout  the following 
expressions, until distinction of the two cases becomes necessary.

From the group law (\ref{law}), two sets of generators can be derived, 
the right- and left-invariant
vector fields, out of which the physical operators of the theory 
and the polarization conditions, 
required to make the quantum representation irreducible,
must be respectively constructed. 

To save unnecessary calculations we shall 
discard the Lorentz subgroup,
which plays no dynamical role, and simply comment on the 
unessential differences
that could be obtained should we keep this subgroup. 
Under this simplification
the two sets of generators are:
\begin{eqnarray}
\XR_{\zeta}&=&i\zeta\frac{\partial}{\partial\zeta}\equiv\Xi \nn \\
\XR_{a_{\mu}(k)}&=
&e^{-ikx}\left[\frac{\delta\;}{\delta a_{\mu}(k)}-
\frac{i}{2}[a^{\mu +}(k)
          +i\phi^+(k)k^{\mu}]\Xi\right] \nn \\
\XR_{a_{\mu}^+(k)}&=&e^{ikx}\left[\frac{\delta\;}{\delta
a_{\mu}^+(k)}+\frac{i}{2}[a^{\mu}(k)
          -i\phi(k)k^{\mu}]\Xi\right] \nn \\ 
\XR_{\phi(k)}&=&e^{-ikx}\left[\frac{\delta\;}{\delta\phi(k)}-
\frac{i}{2}[k^2\phi^+(k)
          -ik^{\mu}a_{\mu}^+(k)]\Xi\right]  \label{XR} \\
\XR_{\phi^+(k)}&=&e^{ikx}\left[\frac{\delta\;}{\delta\phi^+(k)}+
\frac{i}{2}[k^2\phi(k)
          +ik^{\mu}a_{\mu}(k)]\Xi\right] \nn \\
\XR_{x^{\mu}}&=&\frac{\partial}{\partial x^{\mu}} \nn \\ 
& & \nn \\
\XL_{\zeta}&=&i\zeta\frac{\partial}{\partial\zeta}\equiv\Xi \nn \\
\XL_{a_{\mu}(k)}&=&\frac{\delta\;}{\delta a_{\mu}(k)}+
\frac{i}{2}[a^{\mu +}(k)
          +i\phi^+(k)k^{\mu}]\Xi \nn \\
\XL_{a_{\mu}^+(k)}&=&\frac{\delta\;}{\delta a_{\mu}^+(k)}-
\frac{i}{2}[a^{\mu}(k)
          -i\phi(k)k^{\mu}]\Xi \nn \\ 
\XL_{\phi(k)}&=&\frac{\delta\;}{\delta\phi(k)}+
\frac{i}{2}[k^2\phi^+(k)
          -ik^{\mu}a_{\mu}^+(k)]\Xi  \label{XL} \\
\XL_{\phi^+(k)}&=&\frac{\delta\;}{\delta\phi^+(k)}-
\frac{i}{2}[k^2\phi(k)
          +ik^{\mu}a_{\mu}(k)]\Xi \nn \\
\XL_{x^{\mu}}&=&\frac{\partial}{\partial x^{\mu}}-
\int{\frac{d^3k}{2k^0}k^{\mu}
        [a_{\nu}(k)\frac{\delta\;}{\delta a_{\nu}(k)}-
a^+_{\nu}(k)\frac{\delta\;}
        {\delta a^+_{\nu}(k)}]} \nn \\
 &+& \int{\frac{d^3k}{2k^0}k^{\mu}[\phi(k)\frac{\delta\;}{\delta \phi(k)}-
\phi^+(k)\frac{\delta\;}{\delta \phi^+(k)}]} \nn 
\end{eqnarray}
 
From the left generators 
we obtain the following non-trivial commutation relations 
describing, in particular, the dynamical 
content of each parameter:
\begin{equation}
\begin{array}{ll}\left[\XL_{x^{\mu}},\XL_{a_{\nu}(k)}\right]=
ik_{\mu}\XL_{a_{\nu}(k)}  & 
\left[\XL_{x^{\mu}},\XL_{a_{\nu}^+(k)}\right] =
 -ik_{\mu}\XL_{a^+_{\nu}(k)}  \\ 
\left[\XL_{x^{\mu}},\XL_{\phi(k)}\right]=
ik_{\mu}\XL_{\phi(k)}  &\left[\XL_{x^{\mu}},\XL_{\phi^+(k)}\right]=
 -ik_{\mu}\XL_{\phi^+(k)} \\ 
\left[\XL_{a^+_{\mu}(k)},\XL_{a_{\nu}(k')}\right]=
 ig^{\mu\nu}\Delta_{kk'}\Xi & \left[\XL_{\phi^+(k)},
\XL_{\phi(k')}\right]=ik^2\Delta_{kk'}\Xi \\ 
\left[\XL_{a_{\mu}(k)},\XL_{\phi^+(k')}\right]=
k^{\mu}\Delta_{kk'}\Xi & \left[\XL_{a^+_{\mu}(k)},
\XL_{\phi(k')}\right]=k^{\mu}\Delta_{kk'}\Xi \; ,
\end{array} \label{conmutadores}
\end{equation}

\ni where $\Delta_{kk'}= 2k^0 \delta^3(k-k')$ 
is the generalized delta function on the positive 
sheet of the mass hyperboloid.
From the commutation relations (\ref{conmutadores}), line 3, 
we observe that the four components of  
$a_{\nu}(k)$ and $a^+_\nu(k)$ are canonically conjugate variables 
($\hbox{det}(g)\neq 0$)
for any value of $k$, whereas the pair 
$\phi(k)$ and $\phi^+(k)$ are canonically 
conjugate out of the photon mass shell only. 
However, we also note that there is a 
piece of the Lie algebra cocycle, line 4 
(corresponding to the group co-cycle 
$\xi_2(g',g)$),
which mixes $a$'s and $\phi$'s. 
When the $U(1)_{loc}$ parameters themselves acquire dynamical
character, i.e. when the ``photon" is off-shell, 
the cocycle can be
diagonalized, thus transferring  dynamical content 
from one of the pairs 
associated with the new electromagnetic coefficients 
to the new $U(1)_{loc}$ parameters . 
In both cases there are four independent field
degrees of freedom in the group $\tG$, 
but the $T$-equivariance condition (the 
constraints) will remove
two degrees from the electromagnetic field 
in the case $k^2=0$, while in the case 
$k^2=m^2$ the $T$-equivariance condition will 
subtract only one degree of freedom, 
leaving three, which designate a massive vector field (see 
Sec. 2.2 for further discussion).

The GAQ formalism then continues finding the 
left-invariant 1-form $\Theta$ (the {\it quantization form}) associated
with the central generator $\Xi$. The differential $d\Theta$ is a 
{\it presymplectic form} and its {\it characteristic module} 
${\hbox{Ker}}d\Theta\cap{\hbox{Ker}}\Theta$, 
generates  a left subalgebra 
${\cal G}_{\Theta}$ called {\it characteristic subalgebra} 
(the kernel of the Lie algebra cocycle). The quotient 
$(\tG,\Theta)/{\cal G}_{\Theta}$ is a 
{\it quantum manifold}. For our case, they  
prove to be:
\begin{eqnarray}
\Theta &=& \frac{i}{2}\int\frac{d^3k}{2k^0}
\left\{g^{\mu\nu}\left([a_{\mu}(k)-ik_\mu\phi(k)]da_{\nu}^+(k)-
[a_{\mu}^+(k)+ik_{\mu}\phi^+(k)]da_{\nu}(k)]\right.\right. \nn \\
& &+\left.\left. k_\mu[k_\nu\phi(k) +
ia_\nu(k)]d\phi^+(k)-k_\mu[k_\nu\phi^+(k)-
ia^+_{\nu}(k)]d\phi(k)\right.\right. \nn \\  
& &+ \left.\left.[a_\mu(k)-ik_\mu\phi(k)]
[a^+_{\nu}(k)+ik_\nu\phi^+(k)]k^\sigma dx_\sigma \right)\right\} 
+\frac{d\zeta}{i\zeta} \nn \\ 
{\cal G}_{\Theta}&=&<\XL_{x^{\mu}},\,\,\XL_{c(k)},\,\,
\XL_{c^+(k)} > 
\;\;\forall k,\label{charmod}
\end{eqnarray}
\ni where we have defined 
\be
\XL_{c(k)}\equiv \XL_{\phi (k)}+
ik_{\mu}\XL_{a_{\mu}(k)},\;\;\XL_{c^+(k)}\equiv
\XL_{\phi^+(k)}-ik_{\mu}\XL_{a_{\mu}^+(k)}.\label{ideal}
\ee

\ni We should note that ${\cal G}_{\Theta}$ would have
included the Lorentz generators had we kept them in the theory. 
The vector fields in the characteristic subalgebra 
represent the generalized 
classical equations of motion. The invariant quantities under the 
above-mentioned generalized equations of motion, are the 
corresponding generalized Noether invariants defined as:
\begin{equation}
F_{g_j}\equiv i_{\tilde{X}^R_{g_j}}\Theta, \;\;\; \forall g_j\in \tilde{G}.
\end{equation}
\ni For our case, they are:
\begin{eqnarray}
F_{\phi(k)}&=&i_{\XR_{\phi(k)}}\Theta=-ie^{-ikx}k^\mu[k_\mu\phi^+(k)-
i a^+_\mu(k)]\equiv -ik^2\phi^+_{(0)}(k)- k^\mu a^+_{(0)\mu} \nn\\
F_{\phi^+(k)}&=&i_{\XR_{\phi^+(k)}}\Theta=ie^{ikx}k^\mu[k_\mu\phi(k)+
ia_\mu(k)]\equiv ik^2\phi_{(0)}(k)-k^\mu a_{(0)\mu} \nn \\
F_{a_\mu(k)}&=&i_{\XR_{a_\mu(k)}}\Theta=e^{-ikx}[k^\mu\phi^+(k)-i
a^{\mu +}(k)]\equiv k^\mu\phi^+_{(0)}(k)-i a^{\mu +}_{(0)}(k) \\
F_{a^+_\mu(k)}&=&i_{\XR_{a^+_\mu(k)}}\Theta=e^{ikx}[k^\mu\phi(k)+
ia^{\mu}(k)]\equiv k^\mu\phi_{(0)}(k)+ ia^{\mu}_{(0)}(k) \nn \\
F_{x^\mu}&=&i_{\XR_{x^\mu}}\Theta=\int{\frac{d^3k}{2k^0}k_\mu 
[a^\nu_{(0)}(k)-ik^\nu\phi_{(0)}(k)]
[a^+_{(0)\nu}(k)+ik_\nu\phi^+_{(0)}(k)]}, \nn 
\end{eqnarray}
\ni where $a^+_{(0)\mu}(k),\;a_{(0)\mu}(k),\;\phi^+_{(0)}(k),\;\phi_{(0)}(k)$ 
are the initial conditions. 

Apart from the conventional evolution, generated
by $\XL_{x^{\mu}}$, the other two vectors $\XL_{c(k)},\XL_{c^+(k)}$ 
in ${\cal G}_{\Theta}$ (see Eq.(\ref{charmod})) should be understood 
as establishing the equivalence condition:
\begin{eqnarray}
a_{\mu}(k)&\sim&a_{\mu}(k)+ik_{\mu}c(k), \;\;\;
\phi(k)\sim \phi(k)+c(k) \nn \\
a_{\mu}^+(k)&\sim&a_{\mu}^+(k)-ik_{\mu}c^+(k), \;\;\;
\phi^+(k)\sim  \phi^+(k)+c^+(k) \; ,\label{puregauge}
\end{eqnarray}
where $c(k)$ and $c^+(k)$ are the 
corresponding integration parameters. The flow of the vector fields in Eq. 
(\ref{ideal}) constitute the 
{\it gauge transformations} in the theory, and the 
relations (\ref{puregauge}) state simply that pure
gauges do not contribute to the 
symplectic form $d\Theta/{\cal G}_{\Theta}$. The set of vector fields 
(\ref{ideal}) is an ideal of ${\cal G}_{\Theta}$ 
and a horizontal (excluding the
vertical $U(1)$-generator $\Xi$) ideal of
the whole algebra $\tilde{{\cal G}}^L$ of $\tG$. For this subalgebra the 
right-invariant vector fields are proportional to the corresponding left ones  
and, therefore, the Noether invariants are zero. In fact, the proportionality
functions provide a representation of the rest of the generators in the 
subalgebra ${\cal G}_{\Theta}$, more precisely, the Poincar\'e generators. 
These properties characterize the {\it gauge subalgebra}
\be 
{\cal G}_{gauge}\equiv <\XL_{c(k)},\XL_{c^+(k)}>\label{gaugesub}
\ee 
\ni of the theory \cite{gauge}.

Let ${\cal B}(\tG)$ be the set of complex 
valued $T$-equivariant functions $\Psi$ 
on $\tG$, in the sense of principal bundle theory: 
\be
\Psi(t*g)=D(t)\Psi(g), \;\; \forall 
g\in\tG,\;\; \forall t\in T\label{tfuncion}
\ee
\ni where $D$ is a representation of $T$. 
The representation of $\tG$ on ${\cal B}(\tG)$, generated by the 
right-invariant vector fields $\tilde{{\cal G}}^R=\{\XR\}$, is reducible. 
The reduction is achieved by means of the 
restriction imposed by a {\it full polarization} ${\cal P}$, that is, a maximal 
horizontal left subalgebra of $\tilde{{\cal G}}^L$ which contains 
the entire subalgebra ${\cal G}_{\Theta}$. This definition generalizes the 
analogous concept in Geometric Quantization 
\cite{Kostant} where no characteristic 
module exists (since all variables are symplectic). For the group $\tG$, two 
possible polarization subalgebras, corresponding with two possible 
representations, can be chosen  depending on the value of $k^2$. From now on 
we shall distinguish between the cases $k^2=0$ 
and $k^2=m^2\not=0$, placing each in a subsection. The former will lead to the 
Quantum Theory of Photons and the latter to the Quantum Proca Field, both 
requiring the imposition of constraints on 
the wave functions through $T$-equivariant conditions.

\subsection{$\tilde{G}(k^2=0)$ : Electromagnetic Field}

Firstly, we shall consider the case $k^2=0$ 
(null mass\footnote{For this case the whole 
conformal group could be introduced, and
the leaving the mass shell by the photon might be 
interpreted as a breaking 
of the conformal symmetry.}). 
For this case, the polarization subalgebra is:

\begin{equation}
{\cal P}=<\XL_{x^{\mu}},\,\XL_{c(k)},\,\XL_{c^+(k)},\, 
\XL_{a_{\mu}(k)} > \;\;\forall k           
\end{equation} 

As we have already mentioned, the wave functions in GAQ are the complex 
$T$-equivariant functions on the quantization 
group that are nullified by the (left) generators 
in the polarization. We shall
perform the $U(1)$ part of the $T$-equivariance condition, 
impose the polarization 
equations and, separately, write the rest of the $T$-equivariance
conditions, which then will appear as constraints, 
requiring further comments. The
polarized $U(1)$-functions thus satisfy:

\begin{equation}
\Psi(\zeta*g)=\zeta\Psi(g)\;\;\forall g\in\tG, \;\;\;\XL\Psi=0, 
\;\;\forall \XL \in {\cal P},
\end{equation}

\ni or more explicitly:

\begin{eqnarray}
\Xi\Psi&=&i\Psi \nn \\
\frac{\partial \Psi}{\partial x^{\mu}}-i\int{\frac{d^3k}{2k^0}k^{\mu}
[a_{\nu}(k)\frac{\delta\Psi}{\delta a_{\nu}(k)}-a^+_{\nu}(k)\frac{\delta\Psi}
{\delta a^+_{\nu}(k)}]}-& & \nn \\
i\int{\frac{d^3k}{2k^0}k^{\mu}
[\phi(k)\frac{\delta\Psi}{\delta \phi(k)}-\phi^+(k)\frac{\delta\Psi}
{\delta \phi^+(k)}]}&=& 0\nn \\
k_\mu\frac{\delta\Psi}{\delta a_\mu(k)}-
i\frac{\delta\Psi}{\delta\phi(k)}&=&0 \\
k_\mu\frac{\delta\Psi}{\delta a_{\mu}^+(k)}+i\frac{\delta\Psi}
{\delta\phi^+(k)}&=&0 \nn \\
\frac{\delta\Psi}{\delta a_{\mu}(k)}-\frac{1}{2}[a^{+\mu}(k)
          +i\phi^+ (k)k^{\mu}]\Psi &=& 0 \nn 
\end{eqnarray}

\ni with general solution:

\begin{eqnarray}
\Psi(x^\mu,a_{\mu}, a_{\mu}^+,\phi ,\phi^+,\zeta)&=&\zeta 
\exp \left\{\frac{1}{2}\int\frac{d^3k'}{2{k'}^0}
\left(a_{\nu}^+(k')a^{\nu}(k')-
i{k'}^\nu a_{\nu}^+(k')\phi (k') \right.\right.\nn \\
&+& \left.\left. i{k'}^\nu a_{\nu}(k')\phi^+(k')\right)\right\}
   \Phi([a_{\mu}^+ (k)+ik_\mu\phi^+ (k)]e^{-ikx}) \nn \\
  &\equiv & W\cdot\Phi ([a_{\mu}^+ (k)+ik_\mu\phi^+ 
(k)]e^{-ikx}) \label{wavepol}
\end{eqnarray}
where $\Phi$ is an arbitrary power series on the argument 
$[a_{\mu}^+ (k)+ik_\mu\phi^+ (k)]e^{-ikx}$.
The zero-order wave function, i.e. {\it the vacuum}, 
and the one-particle states of  
momentum $k$ are 
$\vert 0>\equiv W$ and
$\vert a_{\mu}^+ (k)>\equiv W\cdot[a_{\mu}^+ (k)+ik_\mu\phi^+ 
(k)]e^{-ikx}$, respectively.

Note that the Gaussian defining the vacuum contains the positive exponent 
$a_{0}^+(k')a^{0}(k')$ which could make the scalar product divergent. This 
breakdown will be avoided after the $T$-function conditions has been imposed 
to define the {\it physical states} of the theory. 
Let us explicitly show how the $T$-function conditions apply for this case. 
According to general settings (see ref. \cite{Ramirez}), the structure group 
$T$ of the group $\tG$ is associated with constraints in the theory. They are 
imposed on the wave functions from the left as in eq. (\ref{tfuncion}) or, 
in infinitesimal form:
\begin{equation}
\XR\Psi_{phys}=dD(\XR)\Psi_{phys}
 \,\,\,\,\forall\, \XR\in {\cal T}\equiv
<\Xi,\XR_{\phi(k)},\XR_{\phi^+(k)}>
\end{equation}
\ni where $dD$ is the ``differential" of the group representation $D$ in 
(\ref{tfuncion}) characterizing the representation. 
We have already imposed the $U(1)$ part, i.e. that corresponding to the 
$U(1)$-function condition $\Xi\Psi=i\Psi$. 
For the rest, we take the trivial representation $dD=0$, and  
we consider the action of $\XR_{\phi(k)}$ and $\XR_{\phi^+(k)}$
on an arbitrary combination of one-particle states 
$\epsilon^\mu (k)\vert a_{\mu}^+(k)>$. Then we get:
\begin{eqnarray}
\XR_{\phi^+(k)}\epsilon^\mu (k)\vert a_{\mu}^+(k)>&=&0 
\;\;\Rightarrow \epsilon^\mu (k)k_\mu=0 \nn \\
\XR_{\phi(k)}\Psi_{phys}&=&0\;\;\Rightarrow 
k^\mu a^+_\mu(k)\Psi_{phys}=0.\label{phystate}
\end{eqnarray}
\ni The first condition establishes that physical  
states must contain the same amount
of longitudinal photons as time-like ones for all $k$. This condition also 
guarantees that physical states 
have positive (or null) norm, since 
$-\epsilon^{\mu +} (k)\epsilon_\mu (k)\geq 0$, and it recovers   
the well-known Gupta-Bleuler condition \cite{Gupta,Bleuler}. The second 
condition in (\ref{phystate}) eliminates the {\it null vectors} (null-norm 
vectors) from the theory. It simply states that all the zero-norm vectors, 
created by the action of $\XR_{\phi(k)}$, are equivalent to zero. In other 
words, the physical wave functions $\Psi_{phys}$ have 
support only on the ``surface" $k^\mu a^+_\mu(k)=0$, thus avoiding the 
above-mentioned breakdown in the scalar product due to the positive exponent 
$a_{0}^+(k')a^{0}(k')$ in the Gaussian 
defining the vacuum. That is, one has to 
integrate only on physical degrees of 
freedom (transversal components) in 
calculating a scalar product. In the 
standard approach (see e.g. \cite{Itzyk}), 
it is usally preferred to allow the null 
vectors to circulate in the system or to 
introduce an equivalence relation 
(``two states differing in a null vector are equivalents") 
and select an equivalence class. This way of proceeding is justified 
by the fact that the mean values of {\it physical operators} prove to be 
independent of the chosen equivalence class  
(see the Sect. 3 for a thorough 
discussion of these facts).

Since the $T$-equivariance condition is 
imposed by means of right generators, it is 
obvious that not all the operators (right 
generators) will preserve the space of
$T$-equivariant states (physical states). 
In the present case the  
operators (named {\it good} in the general approach, 
see Ref. \cite{Ramirez})  preserving this space 
are:

\be
{\cal G}_{good}=<\epsilon_\mu (k)\XR_{a_{\mu}(k)},\,\,
\epsilon^{+}_\mu (k)\XR_{a_{\mu}^+(k)}, \,\,\XR_{x^{\mu}}> \forall k 
\label{good1} 
\ee
where the factors $\epsilon^\mu (k)$ are restricted by the two conditions 
(\ref{phystate}) that is, they verify $\epsilon^\mu (k)k_\mu=0$ and 
$-\epsilon^{\mu +} (k)\epsilon_\mu (k)> 0$. Thus, they project on transversal 
components and keep two field degrees of freedom out of 
the original four field 
degrees of freedom. The {\it good} operators 
in (\ref{good1}) behave as the creation 
and annihilation operators of transversal states:
\be
\epsilon_\mu (k)\hat{a}^{+\mu}(k)
\equiv\epsilon_\mu (k)\XR_{a_{\mu}(k)}, \,\,
\epsilon^{+}_\mu (k)\hat{a}^{\mu}(k)
\equiv\epsilon^{+}_\mu (k)\XR_{a_{\mu}^+(k)} \, ,
\ee
\ni respectively, and the Poincar\'e operators,
\be
P_\mu\equiv i\XR_{x^{\mu}},\,\, M_{\mu\nu}\equiv i\XR_{\Lambda^{\mu\nu}} \; ,
\ee
\ni (when the Lorentz transformations 
$\Lambda$ are kept) declare, in particular, 
that the electromagnetic field carries helicity $\pm 1$. They close 
a Lie subalgebra of the original one, and constitute the {\it physical}  
operators of the theory.

\subsection{$\tilde{G}(k^2\not=0)$ : Proca Field}

As mentioned above, a remarkable characteristic 
of the quantizing group (\ref{law}) is that it 
accomplishes the quantization of both the electromagnetic and Proca fields in 
a unified way. The term in the cocycle proportional to $k^2$ causes the photon 
to acquire mass at the same time that it breaks the conformal invariance of the 
theory. To see this, let us show how it is possible, for this case, to 
decouple the gauge field by means of a transformation which diagonalizes the 
cocycle. In fact, the combinations:

\begin{equation}
\XL_{b_{\mu}(k)}\equiv 
\XL_{a_{\mu}(k)}-i\frac{k^\mu}{k^2}\XL_{\phi(k)},\,\,\,\,
\XL_{b_{\mu}^+(k)}\equiv \XL_{a_{\mu}^+(k)}+
i\frac{k^\mu}{k^2}\XL_{\phi^+(k)}\label{XLb}
\end{equation}
together with $\XL_{\phi(k)}$ and 
$\XL_{\phi^+(k)}$ close the  Lie algebra: 

\begin{equation}
\begin{array}{ll} \left[\XL_{b^+_{\mu}(k)},
\XL_{b_{\nu}(k')}\right]=
iM^{\mu\nu}(k)\Delta_{kk'}\Xi, & \left[\XL_{\phi^+(k)},
\XL_{\phi(k')}\right]=ik^2\Delta_{kk'}\Xi   \\ 
\left[\XL_{b_{\mu}(k)},\XL_{\phi (k'),\phi^+(k')}\right]=0, & 
\left[\XL_{b^+_{\mu}(k)},\XL_{\phi (k'),
\phi^+(k')}\right]=0 \end{array}\end{equation}
where $M^{\mu\nu}(k)\equiv g^{\mu\nu}-\frac{k^\mu k^\nu}{k^2}$. 
The two first commutators are the Proca-like 
and the real Klein Gordon-like ones
respectively, and the others 
simply state that those two fields are decoupled.

For this case, the polarization subalgebra 
is made of the following left generators:

\begin{equation}
{\cal P}=<\XL_{x^{\mu}},\;\XL_{c(k)},\;
\XL_{c^+(k)},\; \XL_{b_{\mu}(k)},\;\XL_{\phi^+(k)} > \;\;\forall k           
\end{equation} 

The integration of the polarization conditions 
essentially follows that of  
Ref. \cite{Garcia}, and, together with the $U(1)$-function condition, 
leads to:

\begin{eqnarray}
\Psi&=&\zeta \exp\left\{-\frac{1}{2}\int\frac{d^3k'}{2{k'}^0}
\left[\sum_{i=1}^3
\alpha^i(k')\alpha^{+i}(k')+
k^2\chi(k')\chi^+(k')\right]\right\}\cdot \nn \\
& &\Phi([\alpha^{+j}(k)e^{-ikx}],[\chi(k)e^{ikx}])\equiv W\cdot \Phi 
\label{wproca} 
\end{eqnarray}
where $\Phi$ is an arbitrary function of its arguments; we have defined 
$\chi(k)\equiv\phi(k)+
i\frac{k^\mu}{k^2}a_{\mu}(k)$, and the coefficients  
$\alpha^i(k),\;\alpha^{+i}(k)$ are the transverse part of 

\begin{equation}
a_\mu(k)=\sum_{\beta=0}^3 \alpha^\beta 
(k)\epsilon_{\mu}^\beta (k),\;\;\;\;a_{\mu}^+(k)=
\sum_{\beta=0}^3 \alpha^{+\beta}(k)\epsilon_{\mu}^\beta(k) \; ,
\end{equation}
$\epsilon_{\mu}^\beta (k)$ being a tetrad defined by

\begin{equation}
\begin{array}{ll} g^{\mu\nu}\epsilon_{\mu}^\beta 
(k)\epsilon_{\nu}^\sigma (k)=
g^{\beta\sigma} & k^\mu\epsilon_{\mu}^i (k)=0,\;\;\; i=1,2,3 \\ 
\epsilon_{\mu}^0 (k)=
k_\mu /k & \sum_{i=1}^3\epsilon_{\mu}^i 
(k)\epsilon_{\nu}^i (k)=
-M_{\mu\nu}(k)  \; . \end{array} 
\end{equation}

Note that we have four field degrees of freedom. For this case, 
the structure group $T$ is itself a central extension and, according to 
general settings (see Ref. \cite{Ramirez}), only a subgroup $T_B=T_p\cup U(1)$  
of $T$ can be consistently imposed as $T$-equivariant condition. 
$T_p$ is a polarization 
subgroup of $T$ which we can choose to be $T_p=<\XR_{\phi(k)}>$. We have 
already imposed the $U(1)$-function condition, so that the rest of the 
$T_B$-function condition states that
\begin{equation}
\XR_{\phi(k)}\Psi=0 \;\Rightarrow \frac{\delta\Phi}{\delta\chi(k)}=0\;\; 
\forall k,
\end{equation}
\ni that is, the arbitrary function $\Phi$ does not depend on 
the $\chi$ variable. On the other hand, if 
we chose $T_p=<\XR_{\phi^+(k)}>$ as the 
polarization subgroup of $T$, we would obtain:
\be
\XR_{\phi^+(k)}\Psi=0 \;\Rightarrow \chi(k)\Psi=0 \; \forall k \; ,
\ee
\ni which simply states that the wave function 
$\Psi$ has support only at the values  
\begin{equation}
\chi(k)=0 \Rightarrow k\phi(k)=-i\alpha^{0}(k) 
\;\;\forall k.\end{equation}
\ni This vaguely resembles  the Higgs Mechanism, where the Goldston bosons  
are eliminated from the theory by making use of the gauge freedom.

In any case, the rest of the wave function, i.e.
\begin{eqnarray}
\Psi= W\cdot \Phi([\alpha^{+j}(k)e^{-ikx}]) \; ,
\end{eqnarray}
\ni is exactly the Proca quantum wave function 
(see \cite{Garcia} for 
more details). The non-trivial good operators for this case are:
\begin{equation}
{\cal G}_{good}=<\epsilon_{\mu}^j (k)\XR_{b_{\mu}(k)},
\,\,\epsilon_{\mu}^j(k)
\XR_{b_{\mu}^+(k)},\,\,\XR_{x^{\mu}}> 
,\;\; j=1,2,3\end{equation}
where $\hat{\alpha}^{+j}(k)\equiv\epsilon_{\mu}^j (k)\XR_{b_{\mu}(k)}$ and 
$\hat{\alpha}^{j}(k)\equiv\epsilon_{\mu}^j (k)\XR_{b^+_{\mu}(k)}$ are the 
creation and annihilation operators of transversal components, respectively. 
The whole Poincar\'e subgroup is also good. The longitudinal components 
$\epsilon_{\mu}^0 (k)\XR_{b_{\mu}(k)}$ and 
$\epsilon_{\mu}^0(k)\XR_{b_{\mu}^+(k)}$ prove to be identically zero on 
polarized functions (\ref{wproca}); even more, they are respectively 
proportional to the {\it gauge} generators $\XL_{c(k)},\;\XL_{c^+(k)}$ and, 
therefore, they also have a null Noether invariant.

\section{Comments and outlook}

The main achievement of the present paper is the quantization of both  the 
electromagnetic and Proca fields within a unified, general scheme  of 
quantization  based on a group structure, which is specially suited for 
dealing with constrained systems. To achieve this goal, the concept of 
gauge symmetry for the electromagnetic field  has been revisited, giving 
rise to a subtle distinction between the constraint and the gauge subgroups. 
Gauge symmetries are associated with horizontal ideals of the general 
symmetry group, the right generators of which prove to have null Noether 
invariants and are, therefore, proportional to the corresponding 
left generators. On the other hand, the constraint subgroup is essentially the 
structure group  of the principal fibre 
bundle $\tG$, which is used as the starting 
point for our Group Approach to Quantization. 

One striking result of the present mechanism is the persistence of the 
gauge symmetry  in the massive case. In this case the possibility 
exists also of not imposing the constraints, 
thus allowing for one extra (scalar) 
massive field with ``negative energy" and  decoupled from the Proca field. 
For the non-abelian theory 
the situation seems to be less trivial and the connection with some sort 
of symmetry-breaking mechanism deserves a further study.

With respect to constraints in the specific $k^2=0$ case, we wish 
to point out that, in principle, and from an algebraic viewpoint, both 
generators $\XR_{\phi(k)}$ and $\XR_{\phi^+(k)}$ can be compatibily imposed 
since, unlike in the $k^2\not=0$ case, their commutator is zero.  It is 
thoroughly argued in the literature that such a set of constraints cannot 
be simultaneously imposed because otherwise the commutator 
$\left[k^\mu \hat{a}^+_\mu(k), \hat{a}_\nu(k')\right]=k_\nu\Delta_{kk'}$ should 
imply  that the state $\hat{a}_\nu(k')\vert \hbox{physical}>$ is 
not in general a physical state; even more, 
not even the vacuum could  satisfy the 
identity $k^\mu \hat{a}^+_\mu(k)\vert 0>=0$. However, in our scheme, 
only a subgroup of operators, 
${\cal G}_{good}$, can be consistently quantized, in fact, the 
transversal components of the electromagnetic field and the Poincar\'e 
generators. In this way, our theory is free from ghosts and null vectors. 
Note that, when the photon is off shell, only one of the two 
$\XR_{\phi(k)}$ or $\XR_{\phi^+(k)}$ operators 
can be consistently imposed as constraint; 
this ``obstruction" has now an algebraic origin: a couple of canonically 
conjugated variables cannot be simultaneously set to zero.

For completion, and in an attempt to extend this formalism to
non-abelian gauge theories, we express the group law (\ref{law}) 
directly in configuration space (see \cite{config}), where 
no knowledge of the solutions of the classical equations of motion is required 
to obtain  non-trivial conclusions about the corresponding quantum theory. 
Although the quantization of the non-abelian gauge 
theories from this new approach is beyond the objectives of this paper, the 
writing of the group law (\ref{law}) in configuration space  
may shed some new light on 
the non-abelian case, which deserves further study. 

The abovementioned group law is 
(we discard, for simplicity, the Poincar\'e subgroup):

\begin{eqnarray}
A''_\mu (x)&=&A'_\mu(x)+ A_\mu (x) \nn \\
\varphi''(x)&=&\varphi'(x)+\varphi(x)\nn \\
\zeta''&=&\zeta'\zeta e^{\frac{i}{2}\int_\Sigma d\sigma_\mu(x) 
J^\mu (g',g)(x)}\nn \\
J^\mu(g',g)(x)&=&A'_\rho(x)\overpmup A^\rho(x) 
+\varphi'(x)\overpmup (\partial_\rho A^\rho(x)) 
-\varphi(x)\overpmup (\partial_\rho A'^\rho(x))\nonumber\\
&&+m^2\varphi'(x)\overpmup \varphi(x) \label{lawconf} 
\end{eqnarray} 
where $\varphi'(x)\overpmup \varphi(x)
\equiv \varphi'(x)\partial^\mu
\varphi(x) -\varphi(x)\partial^\mu \varphi'(x)$ and so on. 
The parameters $A_\mu (x)$ and $\varphi(x)$ have  
to satisfy the equations 

\begin{equation}\begin{array}{rcl}
\partial_\mu F^{\mu\nu}(x) + 
\partial^\nu(\partial_\rho A^\rho(x)) + m^2 A^\nu(x)
\equiv\left[\partial_\mu \partial^\mu + 
m^2\right] A^\nu(x)&=& 0\\ 
\left[\partial_\mu \partial^\mu +
m^2\right]\varphi(x)&=&0\end{array}
\label{eqofmot}\end{equation}
for the current $J^\mu(g',g)(x)$ to 
be conserved ($\partial_\mu J^\mu=0$), 
so that the integral defining the cocycle does not depend on the
chosen space-like hypersurface $\Sigma$  
(see \cite{config} for further details). 

As already mentioned, the 
group law (\ref{lawconf}) suggests a revision of the concept of  gauge 
transformation for the vector potential $A_\mu(x)$. According 
to our scheme, the action of the $U(1)_{loc}$ subgroup in the group law 
(\ref{lawconf}), for $m=0$, leaves 
the vector potential unchanged, although it changes the central parameter 
$\zeta$ (and, accordingly, the phase of the wave function). 
More explicity:
\be
\varphi(x)\rightarrow \varphi(x)+\varphi'(x),\;\;
A_\mu(x)\rightarrow A_\mu(x),\;\;\zeta\rightarrow \zeta 
e^{\frac{i}{2}\int_\Sigma d\sigma_\mu(x) 
\varphi'(x)\overpmup (\partial_\rho A^\rho(x))}\, . 
\ee
\ni However, the standard transformation  (\ref{gauge}) is regained 
as the trajectories associated with the vector fields in the gauge subalgebra 
(\ref{gaugesub}). The same revision applies for the non-abelian case, where 
the situation seems to be a little more subtle.

Finally, we must mention that the analysis in this paper is, in fact, 
a particular case of a more general one in
which a parameter $\lambda$ (see  (\ref{lag2})) 
should be kept arbitrary.

\section*{Acknowledgments}

We all wish to thank J. Guerrero for valuable discussions. 
M. C. and M. N. are 
grateful to the Spanish M.E.C. for a F.P.I. and 
postdoctoral F.P.U. grant, 
respectively, and M. N. to the Imperial College 
for its hospitality.

\pagebreak

\end{document}